%% file: itu-paper.tex
\pdfoutput=1
\documentclass{elsart}
\usepackage{ifpdf}
\usepackage[ascii]{inputenc}
\usepackage[T1]{fontenc}
\usepackage[english]{babel}
\usepackage{amsmath,amssymb,amsfonts,textcomp}
\usepackage{array}
\usepackage{supertabular}
\usepackage{hhline}

\usepackage{subfigure}

\RequirePackage{ifpdf}

\ifpdf
\usepackage[pdftex]{graphicx}
\usepackage[pdftex]{hyperref}
\hypersetup{pdftex, colorlinks=true, linkcolor=blue, citecolor=blue,
  filecolor=blue, pagecolor=blue, urlcolor=blue, pdftitle=, pdfauthor=Kuba
  Moscicki, pdfsubject=, pdfkeywords=}
\else
\usepackage[dvips]{graphicx}
  \usepackage[dvips]{hyperref}
\fi

\newcommand{\VISION}[1]{{}}

\providecommand{\e}[1]{\ensuremath{\times 10^{#1}}}

\usepackage{lineno}

\begin{document}
\begin{frontmatter}
\title{Dependable Distributed Computing for the International Telecommunication Union Regional Radio Conference RRC06}
\author[a:CERN]{Jakub~T.~Mo\'scicki},
\author[a:ITU]{Andrea~Manara \corauthref{cor1}},
\corauth[cor1]{Corresponding author}
\ead{andrea.manara@itu.int}
\author[a:CERN]{Massimo Lamanna},
\author[a:CERN]{Patricia Mendez},
\author[a:CERN]{Adrian Muraru}

\address[a:CERN]{CERN, Geneva, Switzerland}
\address[a:ITU]{International Telecommunication Union, Geneva, Switzerland}

\begin{abstract}

The International Telecommunication Union (ITU) Regional Radio
Conference (RRC06) established in 2006 a new frequency plan for the
introduction of digital broadcasting in European, African, Arab , CIS countries
 and Iran. The preparation of the plan involved
complex calculations under short deadline and required dependable and
efficient computing capability.  The ITU designed and deployed in-situ
a dedicated PC farm, in parallel to the European Organization for
Nuclear Research (CERN) which provided and supported a system based on the
EGEE Grid. The planning cycle at the RRC06 required a periodic execution
in the order of 200,000 short jobs, using several hundreds of CPU hours,
in a period of less than 12 hours.  The nature of the problem required dynamic
workload-balancing and low-latency access to the computing resources.
We present the strategy and key technical choices that delivered a
reliable service to the RRC06.

\end{abstract}
\end{frontmatter}

\linenumbers

\input{intro}

\input{rrc06}

\input{computationalchallenge}

\input{implementation-itu}

\input{implementation-grid}

\input{system-integration}

\input{end}

\end{document}

%% file: intro.tex
\section{Introduction}

\begin{figure}
\begin{center}
\includegraphics[width=100mm]{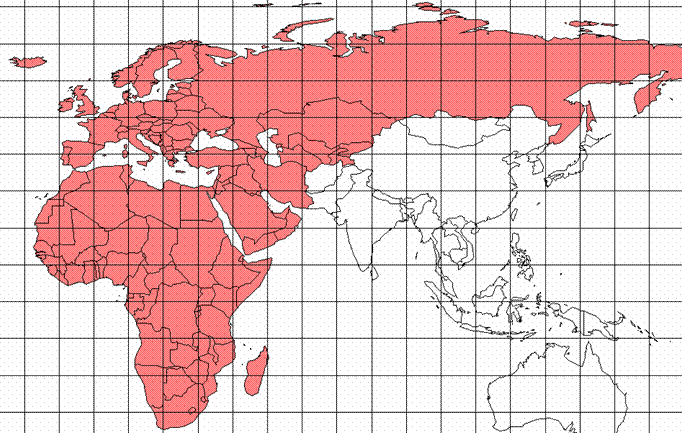}
\label{itu-geographical-extent}
\caption{The extent of the geographical area regulated by the GE06 Agreement.}
\end{center}
\end{figure}

The RRC06 is the second session of the Regional Radiocommunication Conference (RRC) 
for the planning of the digital terrestrial broadcasting service (in band III and IV/V) in European,
African, Arab, CIS countries and Iran(Fig. \ref{itu-geographical-extent}).  
Delegations from 104 Member States of the International Telecommunication Union (ITU \cite{ITU}) gathered in Geneva to negotiate the frequency plan, from the 15th of May to the 15th of June 2006.

The preparation and the organization of this planning conference was
managed by the ITU-R, the Radiocommunication Sector of the ITU.  The
RRC06 Final Acts \cite{RRC06FinalActs} signed by the RRC06
participants constitute a new international agreement, which comprises
the new frequency plan and the procedures for its modification.

Analogue broadcasting has been regulated since 1961
by the Stockholm Agreement in Europe (ST61) and since 1989 by the Geneva Agreement for Africa (GE89). 
The introduction of digital technologies called for a re-planning process in order to optimize
the usage of those frequency bands. The new GE06 plan was designed for
DVB-T (television) and T-DAB (radio) standards, but is flexible enough
to accommodate future developments in digital broadcasting
technologies. 

The technical basis for this planning conference, such as the planning criteria and parameters,
were established in the first session of the RRC ( RRC04 \cite{RRC04}), which was held in Geneva
in May 2004.  During the RRC06 preparatory activities \cite{RRC06PrepAct} it became evident that
one component of the planning process, the compatibility analysis, was
very CPU intensive. The goal of the compatibility analysis is to evaluate the 
interference between broadcasting requirements to identify those that can share the same channel. 
The analysis includes several parameters of the broadcasting requirements such as the geographic location, 
the signal strength and other technical characteristics.

The total capacity required for the compatibility analysis
corresponds to several hundred CPU-days on a high-end 2006 PC. The
compatibility analysis was performed in several iterations. For each iteration
the RRC06 required the output of the compatibility analysis to be
delivered within 12 hours. To support this requirement the
compatibility analysis was split in a large number of parallel
calculations.  The ITU-R implemented a distributed client-server
infrastructure and deployed at its headquarters a dedicated farm
consisting of 84 high-end PCs. A distributed system based on the
EGEE Grid (Enabling Grids for e-ScienE, \cite{EGEE}) and supported by the IT
department of the European Organization for Nuclear Research (CERN)
was deployed, which extended the computing capacity and improved dependability,

The nature of the problem required dynamic workload-balancing and
low-latency access to the computing resources. This fundamental requirement was satisfied both by the ITU system, 
with its dedicated resources, and by the Grid system, by using high-level tools and appropriate customization 
of its infrastructure.

In this paper, we describe in section \ref{sec:RRC06} the RRC06
planning process and in section \ref{sec:ComputationalChallange} the
computational aspects of the compatibility analysis. The implementation of  the 
ITU system is presented in section \ref{sec:ImplementationITU}. The Grid-based system
is analyzed in section \ref{sec:ImplementationGrid} and the integration of the two systems
is discussed in section \ref{sec:Integration}.

%% file: rrc06.tex
\section{The RRC06 planning process}
\label{sec:RRC06}

\VISION{This section contains the background information about the
  RRC06, including the preparatory activities (such as first planning exercise etc)
  but also details the planning process workflow and explains
  different types of the requirements, with the particular emphasis on
  the compatibility analysis.  }

The ITU Constitution\footnote{The ITU Constitution,
  the ITU Convention and the Radio Regulations are the international
  treaties which define the rights and obligations of ITU Member
  States in the domain of the international management of the
  frequency spectrum. } states that ``the radio-frequency spectrum is a
limited natural resource that must be used rationally, efficiently and
economically, in conformity with the provisions of the Radio
Regulations, so that countries or groups of countries may have
equitable access to it''\cite{ITUConstitution}.

The Radio Regulations stipulate  that ``Member States
undertake that in assigning frequencies to stations which are capable
of causing harmful interference to the services rendered by the
stations of another country, such assignments are to be made in
accordance with the Table of Frequency Allocations (where the
frequency blocks are allocated to different radiocommunication services
and to different countries) and other provisions of these
Regulations''\cite{RR4p2}.

\subsection{Frequency Planning}

A frequency plan represents a key mechanism for preserving the rights
of all Member States in the context of equitable access to this
limited resource. Regional Radiocommunication Conferences (RRC) establish
agreements concerning a particular radiocommunication service in specified
frequency bands amongst participating countries.
The last RRC, the RRC06, established the frequency plans (digital and analogue) for terrestrial
broadcasting service (in band III and IV/V) in European, African, Arab,
 CIS countries and Iran. The analogue broadcasting Plan will apply only during the
transition period from analogue to digital broadcasting (up to the 17 June 2015 for most Member States). 
After this period the broadcasting in this band will be regulated only by the digital broadcasting Plan.

Some parts of the frequency bands to be planned at the RRC06 are shared between
broadcasting and other primary services (like fixed and mobile services). The planning process therefore had to
take into account all services which share those bands with equal
rights to operate in an interference-free environment. 

\subsection{The input data}

Member States submitted the input data to the ITU-R in the form of the so-called digital broadcasting requirements. 
The digital broadcasting requirements were
notified as electronic files containing a set of administrative and technical parameters representing the broadcasting
requirements. In addition to the digital broadcasting requirements (about 70K), the
planning process had to take into account assignments to analogue television stations (about 95K) and assignments to other stations (about 10K).  
A fourth type of data, the so-called
administrative declarations (a few million), declared that
incompatibilities between digital broadcasting requirements, analogue television and other services stations may be ignored in the 
frequency synthesis procedure that followed the compatibilities analysis.

Radio communication services are described by administrative and
technical parameters.  For example, administrative parameters include
the notifying administration, site name, geographic location, site
altitude.  Technical paramaters include the power levels, assigned
frequency, network topology, etc.

The digital broadcasting requirements could be submitted at the RRC06 as T-DAB
(radio) or DVB-T (television) standards. Suitable
data elements were provided to accommodate expected development in digital broadcasting technologies.  
Reference Planning Configurations served as simplified models to represent the many system variants (which differ for example in data capacity 
and reception modes) of the requirements. Requirements were submitted as assignments (known
location and transmitter features) or as allotments (only service area
known). Allotments were modeled using
Reference Networks (with different number, location and power of
transmitters) to approximate real networks.

The RRC06 planning approach was based on the protection of service areas for
assignments and allotments and used the statistical model outlined in
the ITU-R Recommendation P1546-1\cite{P1546} to model the
signal propagation.

\subsection{The planning process}

The ITU-R performed two planning exercises after the RRC04 and
prior to the RRC06.  The first planning exercise was run in June
2005 and the second in February 2006. The second planning exercise
established a draft plan which served as input to the RRC06.

The ITU-R and the European Broadcasting Union (EBU)\cite{EBU} developed the RRC06-related software. 
The ITU-R developed the software
for data-capture, data-validation and for the display of the input
data and calculation results, while the EBU developed the planning
software (compatibility analysis, plan synthesis and complementary
analysis). The ITU-R was also responsible for running the planning
software (partly on a distributed infrastructure), producing and
delivering results in due time.

At the RRC06 the frequency plan was established in an iterative way,
as outlined in Fig.2 %\ref{itu-negotiation-workflow}.  
The delegations
engaged in bilateral and multilateral coordination and negotiation
efforts which resulted in a new set of refined digital broadcasting requirements at
the end of every week. Over the weekends the ITU-R
performed the validation of the data and the compatibility analysis and
synthesis calculations. The output of these calculations and the refined frequency
plan were the input for the negotiations in subsequent week, with the last (fourth)
iteration constituting the basis for the final frequency
plan. In order to assist groups of negotiating Member
States, partial calculations were performed for parts of the planning
area in between two global iterations.

The compatibility analysis consisted of the calculation of the
interference between digital broadcasting requirements and other primary services
stations.  For each requirement the compatibility assessment produces
a list of incompatible requirements and a list of available
channels. Three types of compatibility analyses were needed, for both
UHF and VHF frequency bands: digital versus digital (d2dUHF and
d2dVHF), digital versus other services (d2oUHF and d2oVHF) and other
services versus digital (o2dUHF and o2dVHF).

These lists were the input to the plan synthesis process, which
determined a suitable frequency for each requirement in order to avoid
harmful interference and to maximize the number of requirements
satisfied. The RRC06 decided to protect analogue broadcasting services
during the implementation of the digital broadcasting requirements rather than
during the establishment of the plan to maximize the number of requirements satisfied. For this reason each iteration included 
a complementary analysis, which determined which analogue television assignments may suffer
interference from the implementation of a given digital broadcasting assignment or
allotment.

During pre-conference preparatory planning activities only 34\% of
requirements were satisfied. For the first iteration of the RRC06 the
percentage increased to 64\% (UHF) and 74\% (VHF), to reach a
satisfactory 93\% (UHF) and 98\% (VHF) for the final plan.

\begin{figure}
\centering
\includegraphics[width=10cm]{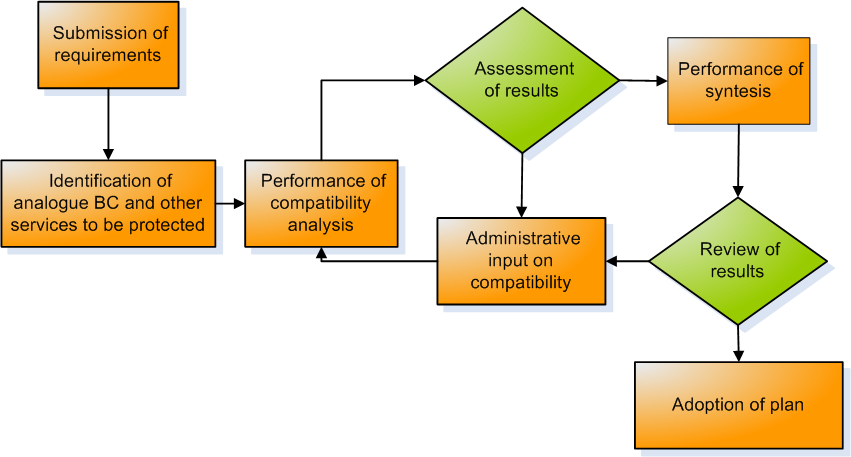}
\label{itu-negotiation-workflow}
\caption{ITU negotiation workflow.}
\end{figure}

%\subsection{Random crap}
%
%
%The first planning exercise was performed by ITU using 6 dedicated 3~GHz PCs plus
%20 1-2 GHz Desktop PCs. The later were only  available overnight and during weekends.
% With this set up more
%than one week was needed to perform the analysis (corresponding to a total 90 days CPU time on
%a 2~GHz PC). The goal of this first planning exercise was for Member States to familiarise with the software and the formulation of their requirements and for the ITU-R to prepare the computing support for the RRC06.
%
%After the first planning exercise ITU-R identified the running time as
%an outstanding issue and decided to take actions by procuring and
%deploying a PC farm within ITU. 
%
%To achieve safety for the RRC06 ITU started the collaboration with CERN eventually ending up with the decision to use grid technologies.
%After a technical kick-off
%meeting at CERN we performed feasibility studies over last two months
%of 2005. The goal was to use the grid to obtain dependable computing power for each compatibility run (taking place on each week-end during the conference) without a-priori reserving CPU resources (CPU resources are in general shared across several user groups). At the same time we wanted to be in the position to run a full compatibility analysis within hours. 

%% file: computationalchallenge.tex
\section{The computational challenge}
\label{sec:ComputationalChallange}

\VISION{This section describes our activity from the computational
  point of view: size of the problem, the deadline, the structure of
  the computation, the time constraints, the nature of static and
  dynamic optimization. It refers to the key elements of the planning
  process described in the previous section: different types of
  requirements, iterative workflow etc.}

The compatibility assessment is CPU-intensive.  In the compatibility
analyses each requirement must be run against all the others, for six
different types of analysis (d2dUHF, d2dVHF, d2oUHF, d2oVHF, o2dUHF,
o2dVHF). In this paper we use the term {\em atomic calculations} to
refer to individual, indivisible calculations defined in compatibility
analysis datasets. The term {\em task} refers a unit of work which
corresponds to a set of atomic calculations. The term {\em job} is
used in the context of Grid job submission only.

For the first planning exercise the atomic calculations were clustered
in tasks of 100 for all types of analyses. With the limited resources
available at that time, that exercise took about one week (elapsed
time), for an integrated 90 CPU days.

\begin{figure}
\centering
\includegraphics[width=13cm]{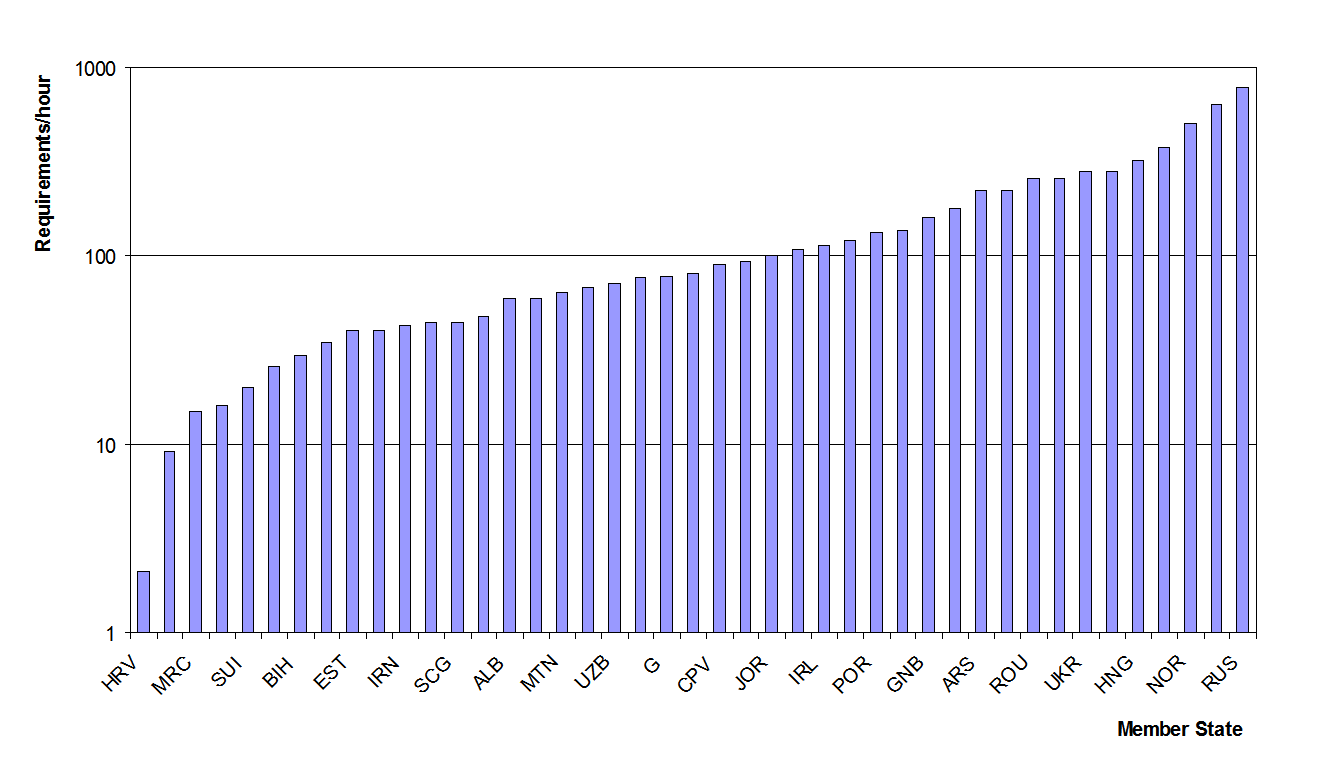}
\caption{Distribution of the number of processed requirements per hour for the d2dUHF analysis as a function of the Member State. Data for the first planning exercise.}
\label{FirstPE_stats_NoReqHours}
\end{figure}

The detailed study revealed an exponential distribution of the
requirement processing time which spans almost three orders of
magnitude (Fig. \ref{FirstPE_stats_NoReqHours}). The huge variation in
running time depends, among other parameters, on the number of acceptable channels specified in
the digital broadcasting requirement, the requirement type (assignment versus
allotment), the network topology and signal propagation zones specific to
the geographical area of the Member State.

Further investigation showed that a complete static optimization of
the load \footnote{The static optimization of the load is an ability
  to a priori cluster the requirements, so that the execution time of
  each cluster is equal.}  was not possible due to the unpredictable
nature of the data as the Member States could change their
requirements before each RRC06 iteration. On the other hand, there was 
clearly a need to create smaller clusters for the most CPU demanding
type of analysis d2dUHF and d2dVHF, minimizing the spread between the
shortest and longest tasks. Table \ref{itu-grouping-table} shows the
granularity chosen for the different types of analysis in the RRC06
iterations for the Grid and ITU systems. The granularity was adjusted
manually in between the iterations. The load balancing was handled
dynamically at runtime.

\begin{table}
\centering
\begin{tabular}{|c|c|c|c|c|c|c|c|}
\hline
iteration & d2dUHF  & d2dVHF  & d2oUHF & d2oVHF & o2dUHF & o2dVHF\\ \hline

 1 & 3(3)  & 5(5)  & 100(100) & 100(100)  & 100(100)  & 100(100)  \\

 2 & 4(3)  & 4(10)  & 50(100)  & 50(100)   & 100(100)  &  100(100) \\

 3 & 2(3)  & 2(5)  & 50(100)  & 50(100)   & 50(100)   &  50(100)  \\

 4 & 2(3)  & 2(10)  & 50(100)  & 50(100)   & 50(100)   &  50(100)  \\ \hline
\end{tabular}
\caption{Compatibility analysis granularity for the RRC06 iterations for Grid and ITU (in parenthesis) system.}
\label{itu-grouping-table}
\end{table}

The workload for each compatibility analysis run at the RRC06
corresponded to some several hundred CPU hours.  Additionally the
workload was to be completed within a deadline of a few hours. The time
constraints were critical: an hypothetical problem with timely
delivery of analysis results could have resulted in a failure of
international negotiations.

%Given the
%cyclic shedule envisaged at the RRC06 the time constraints were
%critical: missing a deadline might have had implied loosing an
%iteration which in turn may have implied less requirements to be
%accomodated at the RRC06. 

%For the Second Planning Exercise and during subsequent negotiation
%phases 

The total CPU demand decreased with each RRC06 iteration. 
Member States decreased the number of
requirements and the number of acceptable channels for each
requirement, reducing therefore the total workload at each analysis
iteration. Finally, as the frequency plan was refined during successful
negotiations between the Member States, the number of conflicting
requirements also decreased. The CPU demands for the ITU and Grid
systems is presented in the next sections.

%% file: implementation-itu.tex
\section{ITU system}
\label{sec:ImplementationITU}

\VISION{this section explaing the implementation and results obtained on the ITU cluster, including
the monitoring and performance data}

\begin{figure}
\centering
\includegraphics[width=10cm]{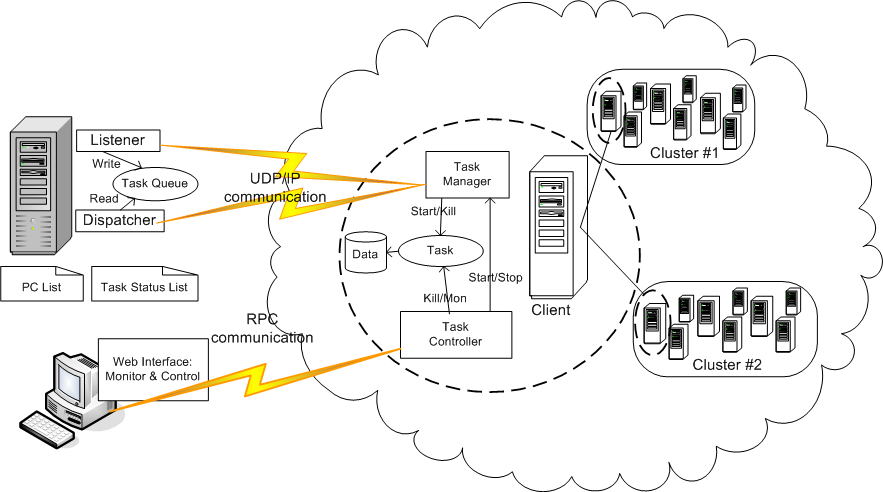}
\label{itu-br-dedicated-system-architecture}
\caption{Architecture of the ITU dedicated system.}
\end{figure}

The ITU system consisted of a client-server distributed system running
on a dedicated PC farm. The farm resources evolved in time. Initially it consisted of six high-end
dedicated PCs complemented by some tens of ITU staff desktop PCs, available only overnight and during weekends. 
Using this configuration, the calculations for the first planning exercise
required about one week, showing that the running time was an outstanding
issue in preparation for the RRC06. The ITU-R therefore decided to buy a PC farm, which was deployed within
ITU headquarters by the  ITU Infrastructure Services department (ITU
IS).In its final configuration at the RRC06 the farm was composed of 84 high-end dedicated 3.6 GHz hyper threading PCs. 
Accurate measurements showed that hyper threading permits 
to gain about 30\% in computing time by running two tasks in parallel on one PC with respect to the situation when the same tasks are run sequentially.

\begin{table}
\centering
\begin{tabular}{|c|c|c|c|c|c|}
\hline
Iteration & $N_{calc}$ & $N_{task}$ &  $t_{total}$ & $t_{clients}$ \\
\hline
 1 & 173K  & 26K  & 5.9h & 621h \\ % 84*2 

 2 & 168K &  23K  & 4.1h & 463h  \\

 3 & 154K &  23K  & 3.4h & 300h  \\

 4 & 155K &  21K  & 2.6h & 205h  \\ \hline

\end{tabular}
\caption{Performance of the ITU system (84*2 simultaneous processes) during compatibily analysis calculations}
\label{itu-system-performance-table}
\end{table}

To cope with redundancy and logistic issues (available space, power and cooling consideration), ITU-IS decided 
to deploy the farm into two separate clusters. 
The first cluster consisted of 47 PCs and was equipped with optical fibers and a 1Gb/s network switch, 
while the second cluster consisted of 37 PCs with a slower 200Mb/s network switch.
This configuration did not significantly impact on the performance of the system.

The architecture layout is presented in
Fig. \ref{itu-br-dedicated-system-architecture}.  The system was
implemented with Perl scripts installed as Windows services and a
custom communication protocol based on UDP/IP.  The UDP packets
carried information on the executable to be run and on the relevant
input parameters.  In the reliable internal network of the ITU farm
the packet loss was not a problem. The server implemented two Windows
services, a Listener and a Dispatcher, responsible for task
submission, task management and workload balancing. To cope with
high-load, the TaskQueue file ensured asynchronous operation of the
system and prevented packet lost. The system automatically managed the
task status and resubmitted the ones which were not completed.

%Configuration files included the
%list of available PCs and the list of jobs to be executed.  

\begin{figure}
\centering
\includegraphics[width=10cm]{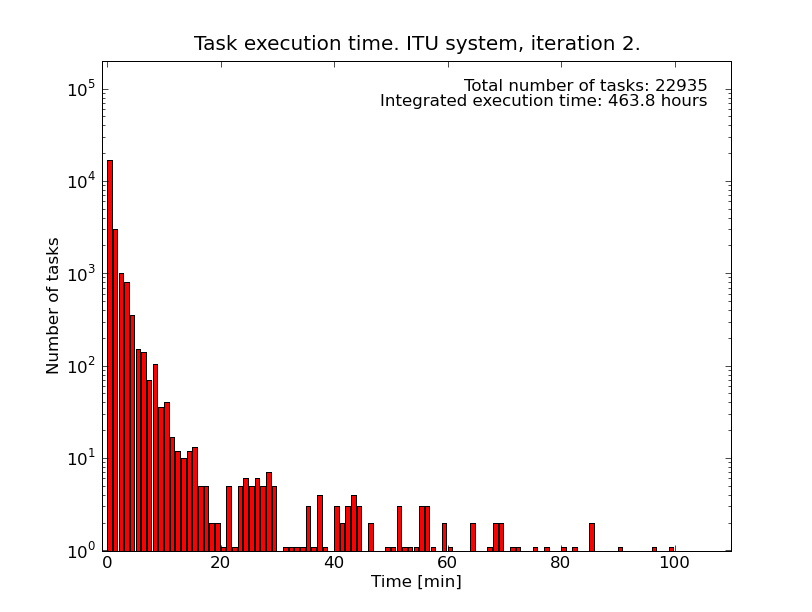}
\label{itu-cluster-job-execution-histogram-iteration-2}
\caption{Distribution of the elapsed time for the ITU system during RRC06 iteration 2.}
\end{figure}

The clients implemented two Windows services, the TaskManager
responsible for running tasks according to Dispatcher requests and the
TaskController responsible for monitoring and control operations. A web
application (implemented with ASP.NET and C\#) running on a
dedicated machine (WebInterface), provided monitoring and control
interfaces to operate the system.

In the first phase, the client installation on non-dedicated resources
(desktop PCs) was implemented using a MSI-compatible installation
procedure managed by Windows Systems Management Server (SMS). In the
dedicated farm, the software and data were deployed on a shared folder
and copied directly to the client PCs.  MD5 checksums were performed
to insure data consistency. At system startup the server automatically triggered the software and data installation at the client.

The system supported 2*84 simultaneous tasks most of the time with negligible job loss.  Software
and data installation involved ~350 MB to be deployed in 2*84 folders
and took on average 15 minutes for the entire farm.

\begin{figure}
\centering
\includegraphics[width=10cm]{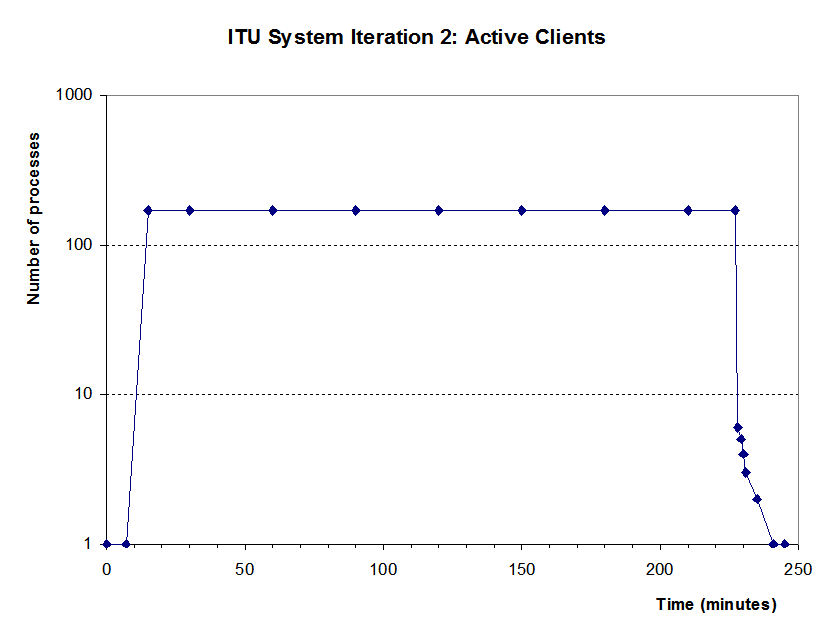}
\label{itu_iter2_evolution}
\caption{Number of running processes as a function of time during RRC06 iteration 2.}
\end{figure}

The performance of the ITU system is reported in Table
\ref{itu-system-performance-table}, where the total workload of atomic calculations $N_{calc}$, the number of tasks $N_{task}$, 
the total time to complete the iteration $t_{total}$ and the integrated elapsed time on the
clients $t_{clients}$ are shown for each iteration. 
The distribution of the tasks processing time for the ITU system during iteration 2 of the RRC06 
is shown in Fig.~5. %\ref{itu-cluster-job-execution-histogram-iteration-2}.
The evolution of the number of running processes as a function of time during RRC06 iteration is shown in Fig.~6.%\ref{itu_iter2_evolution}. 
This last figure illustrates interesting features of the ITU system: the dynamic load balancing (about 96\% of the 
clients complete processing tasks practically at the same time) and limited submission latency (about 15 minutes, the time necessary for the clients 
to download the latest version of software and data at server start-up).

Taking into consideration also the four runs of complementary analysis and the partial runs during multilateral negotiations,
the ITU system at the RRC06 ran more than 180 thousand tasks for an overall integrated elapsed time of 4500 CPU/hours, 
i.e. more than half a CPU year.

%% file: implementation-grid.tex
\section{Grid system}
\label{sec:ImplementationGrid}

\VISION{This section describes the Grid-based solution, including:
the background info on EGEE, architecture built around Ganga/DIANE, 
setting up services for reliable computing/high-availability, operation of the system including 
the installation of software, summary of the runs, description of static optimization performed,
performance analysis (latency,speedup and overheads)}

Enabling Grids for E-sciencE (EGEE) is a globally distributed system
for large-scale batch job processing.  At present it consists of around 300 sites
in 50 countries and offers more than 80 thousand CPU cores and 20 PB
of storage to 10 thousand users around the globe. EGEE is a
multidisciplinary Grid, supporting users in both academia and
business, in many areas of physics, biomedical applications,
theoretical fundamental research and earth sciences.  The largest user
communities come from the High-Energy Physics, and in particular the
experiments active at the CERN Large Hadron Collider (LHC).

The EGEE Grid has been designed and operated for non-interactive
processing of very long jobs.  A set of complex middleware services
integrate computing farms and the batch queues into a single, globally
distributed system. The access to the distributed resources is
typically controlled by the fair-share mechanisms, ensuring usage of
resources by groups of users according to predefined policies.  In
typical configurations a
large number of users share individual computing resources across multiple Virtual Organizations
(VOs)\footnote{Virtual Organization is a group of users sharing the
  same resources. Members of one Virtual Organization may belong to
  different institutions.}  This architecture is suitable for
high-throughput computing but is not efficient for high-performance,
short-deadline, dependable computing which is stipulated by the RRC06
compatibility analysis application.

In the EGEE Grid environment and on a short time-scale these
requirements may only be implemented if high-level tools are used to
control the job workload and the Grid infrastructure is appropriately
customized.

\subsection{The tools}

To run RRC06 compatibility analysis application Ganga and DIANE tools were used.

Ganga provides a uniform and flexible interface to submit, track and
manipulate jobs \cite{Ganga}. DIANE is an agent-based job scheduler which provides
fault-tolerant execution of jobs, dynamic workload-balancing and
reduced overhead in accessing the computational resources \cite{DIANE}.

\begin{figure}
\centering
\includegraphics[width=13cm]{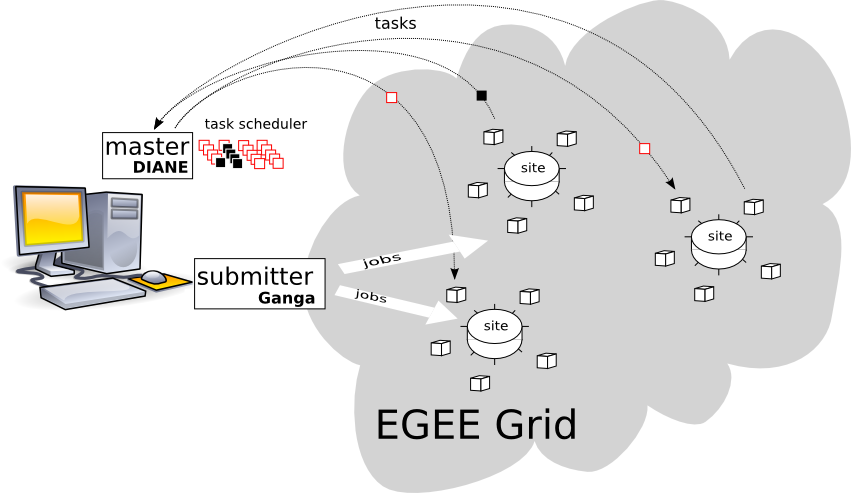}
\caption{Overview of the Grid system based on Ganga/DIANE.}
\label{itu-ganga-diane-grid-overview}
\end{figure}

The outline of the architecture is presented in
Fig.\ref{itu-ganga-diane-grid-overview}.  Worker agents are submitted to
the Grid and pull the tasks from the Master server
which controls the distribution of the workload. The system is
fault-tolerant and may run autonomously: a Worker agent which fails
to complete the assigned calculations is replaced by another Worker
agent. The overhead of scheduling the calculations is negligible in
comparison with the overhead of classic Grid job submission. The
system dynamically reacts to changing workload and provides dynamic
load-balancing. The results of the compatibility analysis of the
requirements are directly uploaded to the Master server.  The
implementation of the RRC06 system on the EGEE Grid was based on DIANE
1.5.0 and Ganga 4.1.

The input data, including the specification of the digital broadcasting requirements
and the tuned compatibility analysis application, were distributed to
the collaborating Grid sites shortly before the analysis was
launched. The 100MB installation package was deployed into the
directory mounted on a shared file system accessible by all worker
nodes of a collaborating Grid site (so called ``software areas''). The
installation was managed by separate grid jobs running with the credentials of the VO manager and using MD5 check-sums to assure consistency of the
installation tarballs. The installation was automated and the
installation jobs checked periodically to download the installation
packages available in a central repository at CERN. This allowed to
automatically distribute the new installation packages in 15 minutes
after the ITU-R made them available.

The ITU personnel updated the software packages with 2 hours' notice. In this time window the grid system had
to be up and ready to start the computation at full speed, as soon as
the update was available.

\subsection{The infrastructure}

The access to the computing resources on the Grid for the RRC06 use
was implemented using the GEAR Virtual Organization (vo.gear.cern.ch).
The CPU demand for RRC06 was much smaller than typical Grid
applications which require huge throughput over very long periods of
time.  However, conversely to many other Grid applications,
availability of resources within well-defined and strict time
constraints was critical. Therefore a number of high-availability centres
in the EGEE Grid \footnote{CERN, CNAF+few other sites(I), PIC(E),
  DESY(D), MSU(RU) , CYFRONET(PL)} were involved. The resources at
these centres were not dedicated to the RRC06 activity, however the
job priority parameters were adjusted during short periods of
intensive processing of the RRC06 compatibility analysis (the weekends
between the major conference iterations). On average 300 CPUs were
observed to be available at all times with occasional peaks of
c.a. 600 CPUs.

Redundant deployment of key services, such as the Master servers, Grid
User Interfaces and Resource Brokers \cite{LCG} allowed for fail-over in case
of problems. For storing the application output the AFS and local filesystem
were used simultaneously.

\subsection{Analysis of the system}

\begin{table}
\centering
\begin{tabular}{|c|c|c|c|c|c|c|c|}
\hline
iteration & $N_{calc}$ & $N_{task}$ & $t_{total}$  & $t_{worker}$ & $N_{worker}$ & $r_{fail}$\\ \hline

 1 & 243K  & 26K  & 6h40m & 425h  & 190  & \textless $3\e{-4}$ \\

 2 & 237K &  23K  & 6h30m & 332h  & 125  &  $4\e{-5}$ \\

 3 & 224K &  40K  & 1h35m & 192h  & 210  &  0 \\

 4 & 218K &  39K  & 1h5m & 151h  & 320  &  0 \\ \hline
\end{tabular}
\caption{Summary of RRC06 compatibility analysis iterations.}
\label{itu-summary-table}
\end{table}

% COMMENT: run 3: 1.35h - all workload finished, the master was
% blocked on worker kill due to network communication problem and
% message architecture (bidirectional) givin total 3.05h

The summary of RRC06 iterations is presented in Table
\ref{itu-summary-table}.  For each analysis iteration the total workload
consisted of $N_{calc}$ atomic calculations. The calculations were
executed in bunches according to previously defined static clustering
(section \ref{sec:ComputationalChallange}). The $N_{task}$ tasks were
distributed dynamically to the $N_{worker}$ Worker agents. The Worker
agents were submitted as {\em jobs} and executed on the Grid worker
nodes. $t_{total}$ is the makespan or the total time to complete the
compatibility analysis.  $t_{worker}$ is the integrated elapsed time on
the worker nodes. $r_{fail}$ is the reliability of the system and
corresponds to the number of failed tasks which could not
automatically recover. With fewer than 10 lost tasks in run 1 and one
lost task in run 2 the reliability of the system exceeded by few orders
of magnitude the reliability of the Grid infrastructure.

\begin{figure}
\centering
\includegraphics[width=10cm]{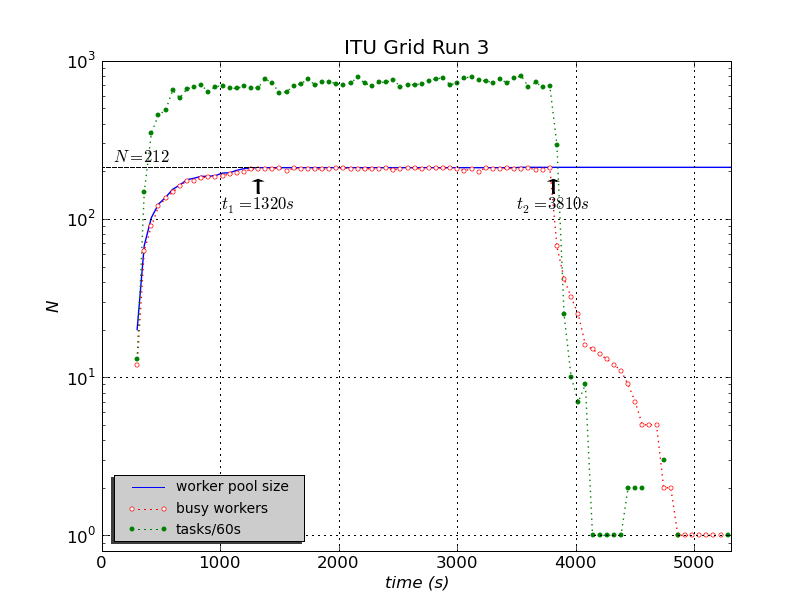}
\caption{Run 3 workload. Resolution=60s. }
\label{itu-run3-history-plot}
\end{figure}

Contrary to the ITU system which used a fixed set of resources, in the
Grid resources are dynamic: a different set of worker nodes is used at
each iteration.  The worker node characteristics such as the CPU and
memory also show large variations. Therefore a direct comparison of
$t_{total}$ and $t_{worker}$ parameters between ITU and Grid runs is
not possible.

%%The input dataset is composed of 6 different types of requirements, as
%%explained in section \ref{sec:ComputationalChallange}.  The average
%%processing time of the requirement depends on the type.  Before the
%%analysis began appropriate task granularity (number of requirements in
%%a bunch) was selected according to the type of the requirements
%%(Fig. \ref{itu-grouping-table}). The goal of this static optimization
%%of the workload was to reduce the number of tasks and minimize the
%%spread of the task execution time.
%%
%\TODO{let{\textquotesingle}s analyze the clustering: how to cluster the
%data sample in such a way that the probability of having at least one
%cluster longer than some specified time (absolute or relative) is more
%than N\% over 4 rounds? From this we could derive this probability for
%the cluster size finally used...
%}

%\TODO{a posteriori analysis: given the clustered tasks (randomized) can
%we somehow see how different were the datasets? How much they mutated?}

\begin{figure}
\centering
\includegraphics[width=10cm]{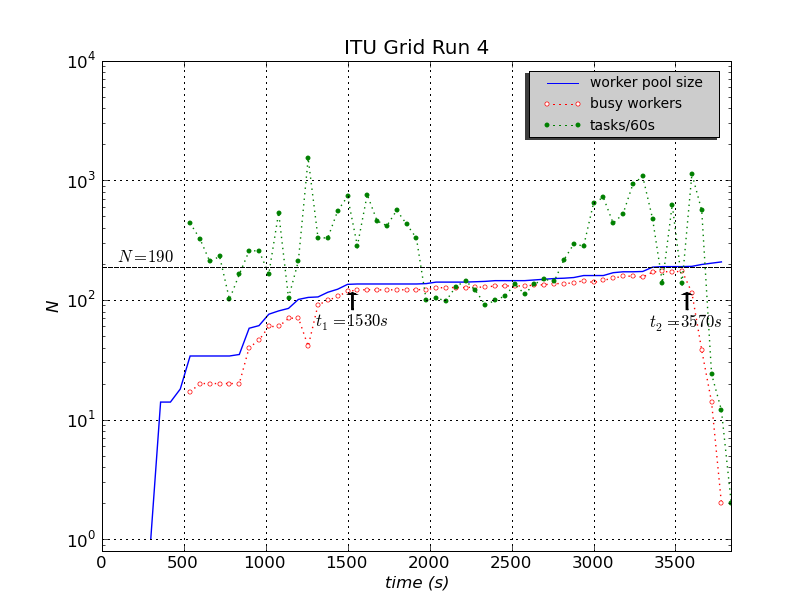}
\caption{Run 4 workload. Resolution=60s. The point $t_1$ was selected arbitrarily. In run 4 two parallel master servers were used and this figure corresponds to one of the masters and half of the total workload.}
\label{itu-run41-history-plot}
\end{figure}

The efficiency of the system depends on the Grid job submission
latency, efficiency of task scheduling and workload balancing.
Fig. \ref{itu-run3-history-plot},\ref{itu-run41-history-plot} show the
workload distribution for selected runs. $N_w$ worker agents are submitted at $t_0=0$.  In the submission
phase, $t<t_1$, the throughput of the system is limited by the
submission latency. As the pool of worker nodes increases the target of
$N_w$ workers is reached at time $t_1$. In the main processing phase,
$t_1<t<t_2$, the pool of worker nodes remains stable and the system
throughput mainly depends on the efficiency of scheduling. At time
$t_2$ the number of remaining tasks becomes smaller than the number of
processors in the pool. In this phase the execution time is dominated
by the workload-balancing effects from few slowest tasks. %makespan

The number of available worker nodes may vary significantly in the
Grid from one run to another. The contribution of the job submission
latency to the total execution time may be approximated by the area
between the target line and the worker pool size curve. In run 3 the
latency of job submission corresponded to 12\% of the total execution
time, whereas in run 4 it corresponded to 48\%: 33\% in the submission
phase and 15\% in the main processing phase. 

\begin{figure}
\begin{minipage}[b]{0.5\linewidth} % A minipage that covers half the page
\centering
\includegraphics[width=6cm]{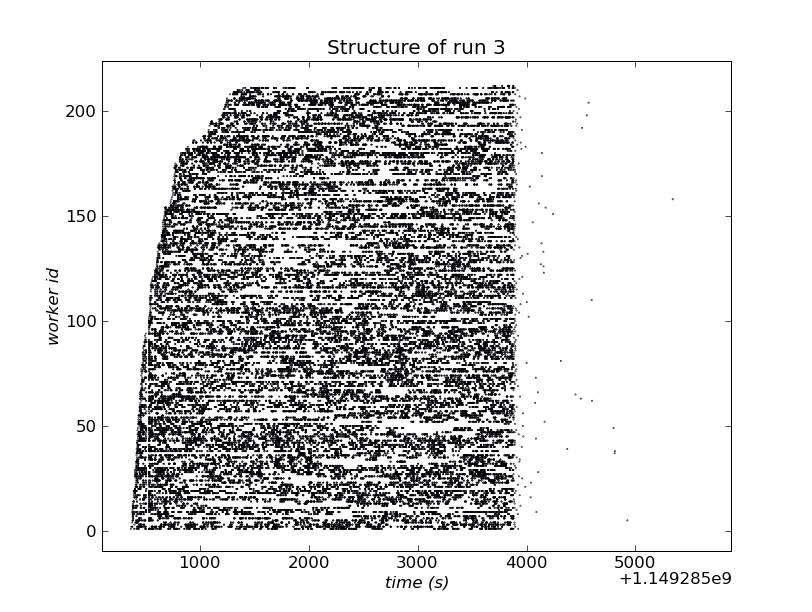}
\caption{Run 3 profile.}
\end{minipage}
\hspace{0.5cm} % To get a little bit of space between the figures
\begin{minipage}[b]{0.5\linewidth}
\centering
\includegraphics[width=6cm]{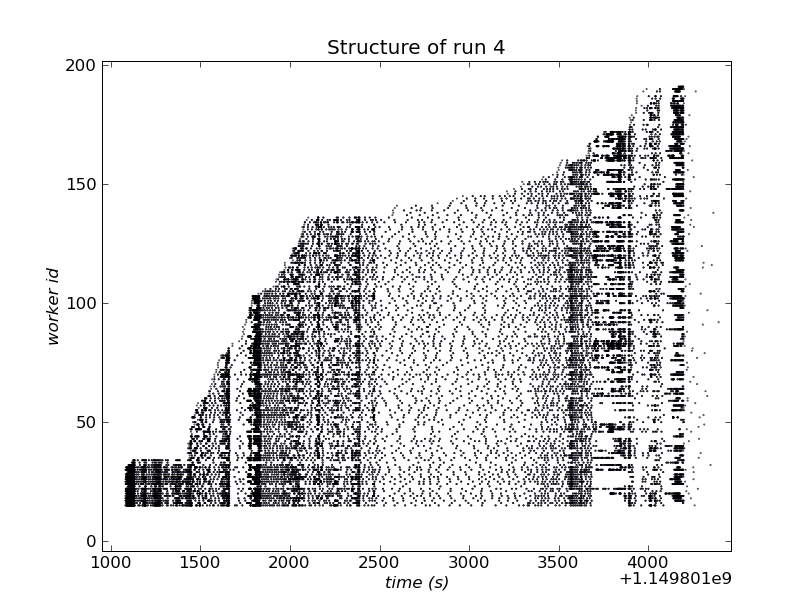}
\caption{Run 4 profile.}
\end{minipage}
\label{itu-run3and41-scatter-plot}
\end{figure}
               
The integrated difference between the worker pool size and the number
of busy workers corresponds to the scheduling overhead. This overhead
includes the network latency and throughput as well as the task
handling efficiency of the master server. In run 3 the scheduling
overhead in the submission and processing phases corresponded to
2-3\%. In run 4 the 30\% scheduling overhead in the submission phase
was observed and 10\% in the processing phase. 

The unbalanced execution of the slowest tasks in the last phase
contributes 26\% of the total execution time in run 3 and to 5\% in
run 4. In this phase the utilization of available resources was very
low, 5\% in run 3 and 20\% in run 4. The majority of the workers in
the pool remained idle while the few remaining tasks were being finished.

%%\begin{figure}
%%\centering
%%\subfigure[run3]
%%{
%%  \label{R3.}
%%  \includegraphics[width=6cm]{itu-run3-scatter}
%%}
%%\hspace{1cm}
%%\subfigure[run41]
%%{
%%  \label{R41.}
%%  \includegraphics[width=6cm]{itu-run41-scatter}
%%}
%%\end{figure}

The striking difference of scheduling and workload-balancing
efficiency between runs 3 and 4 may be explained by the task
scheduling order which reflects the internal input data structure.
The run profile plots are shown in Fig.~10,~11. %\ref{itu-run3and41-scatter-plot}.
%and \ref{itu-run41-scatter-plot}.  
Point (t,w) in the run profile
represents a task completed by worker w at time t. In run 4 the tasks
are drawn directly from the input data in the natural order and
clusters of very short tasks created a very high load on the server. The
long tasks were processed in the middle of the run and did not affect
the overall load-balancing. In run 3 the tasks were selected in a random
order by the scheduler. The momentary load on the server was
reduced. The tasks were scheduled more uniformly across the entire
run. There were a few long tasks at the end of the run that resulted in poor
load-balancing.

The intrinsic job submission latency in the Grid prevents the running
of a large number of short jobs in a short time, unless user-level tools
such as DIANE are used. For RRC06 using DIANE allowed to reduce the
Grid overheads and provided efficient management of a large number of
tasks.  Additionally a runtime workload balancing allowed to evenly
distribute a workload without precise, a priori knowledge of the
task execution times in the dataset. The overhead reduction and
workload balancing were the crucial factors of the successful usage of the Grid
for the RRC06.

%run3:
%time window size = 60s
%zone 0 [0,1320] scheduling efficiency=0.97 resources idle=10.95%
%zone 1 [1320,3810] scheduling efficiency=0.98 resources idle=1.02%
%zone 2 [3810,5310] scheduling efficiency=0.05 resources idle=26.71%

%run41:
%time window size = 60s
%zone 0 [0,1530] scheduling efficiency=0.71 resources idle=33.06%
%zone 1 [1530,3570] scheduling efficiency=0.90 resources idle=14.61%
%zone 2 [3570,3840] scheduling efficiency=0.21 resources idle=4.86%

%\TODO{
%Speedup. Remark 1: as we scale-down the size of the problem from initial
%425 CPUh to 150 CPUh, keeping number of parallel task more or less
%constant (from 234K to 218K) and increasing the number of processors
%(except at iteration 2) the speedup according to the
%Amdhal{\textquotesingle}s low should also decrease. 
%}
%

%%Fig. \ref{itu-runTEST-resource-profile} shows the number of concurrently processed
%%task in the function of time. It should be noted that the efficiency drops when the message rate on the
%%server is higher. This may be observed by comparing with Fig. \ref{itu-runTEST-task-profile}.
%%
%%Fig. \ref{itu-server-message-rate} shows the message rate (receiving
%%and sends out tasks) at the master.  Message rates up to 100Hz may be
%%easily sustained by the master. The bottom line is that DIANE enables
%%to run efficiently very short tasks on the grid.

%%\begin{figure}
%%\centering
%%\includegraphics[width=10cm]{itu-server-message-rate}
%%\caption{TODO: TEST RUN (TO BE REPLACED BY THE REAL RUNS...)}
%%\label{itu-server-message-rate}
%%\end{figure}

%% file: system-integration.tex
\section{System Integration}
\label{sec:Integration}

\VISION{This chapter describes how the systems were integrated. Not sure if we want to put it as a separate chapter or 
  rather to conclusions.}

%The two distributed infrastructure had each one its own monitoring
%software. We were also able to integrate the two infrastructures at
%the monitoring level using the Caltech-developed MonALISA
%architecture (ref. ). The ApMon Perl module was used to send UDP
%packets to the monitoring server.  Fig. \ref{MonaLisa1} and
%Fig. \ref{MonaLisa2} below shows two ITU clusters monitored together
%with several EGEE sites. The possibility to access seamlessy
%resources from grid and corporates (which we exploited only at the
%monitoring level) is very appealing and should be interesting for
%other users. Ideally one should have at his disposal dedicated
%in-situ resources for faster response and grid resources when facing
%peak demand.

The Grid and ITU systems were integrated at the monitoring level using
the MonALISA framework (Monitoring Agents in A Large Integrated
Services Architecture, developed by Caltech University \cite{Monalisa}).
MonALISA provides a set of pluggable distributed services for
monitoring, control, management and global optimization for large
scale distributed systems.

\begin{figure}[h]
\centering
\includegraphics[width=6cm]{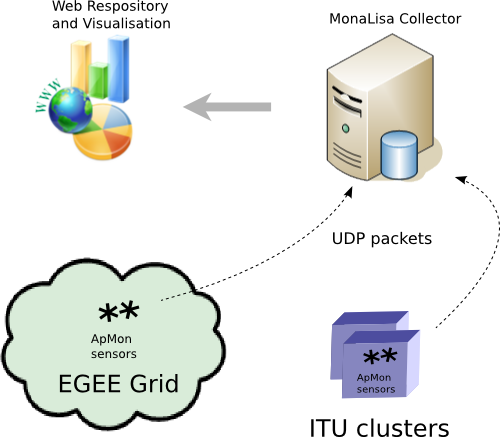}
\label{blah-itu-monalisa-integration-diagram}
\caption{System integration via Mona Lisa monitoring.}
\end{figure}

To collect and combine monitoring information from both ITU
and Grid systems, the following software components were
deployed: instances of MonALISA collector service, web-enabled data
visualization repository and custom ApMon monitoring sensors on worker
nodes (Fig. 12).%\ref{blah-itu-monalisa-integration-diagram}). 
ApMon, the monitoring API,
allows to send fine-grained custom monitoring parameters into the
MonALISA collector service. The ApMon uses UDP datagrams to transport
the XDR-encoded information \cite{XDR} and includes a sequence number to
verify the integrity of all monitoring reports. In addition, ApMon
provides out-of-the-box system monitoring of the host, including usage
of system resources such as memory or CPU.  Monitoring parameters of
ApMon, such as monitoring frequency and collector destination, may be
dynamically configured by remote services. ApMon implementations are
provided for different programming languages, including C, C++, Java,
Perl and Python. The cross-language support has proven to be useful in
the case of RRC06 as the ITU system was built in Perl while the Grid 
used Python.

\begin{figure}
\centering
\includegraphics[width=12cm]{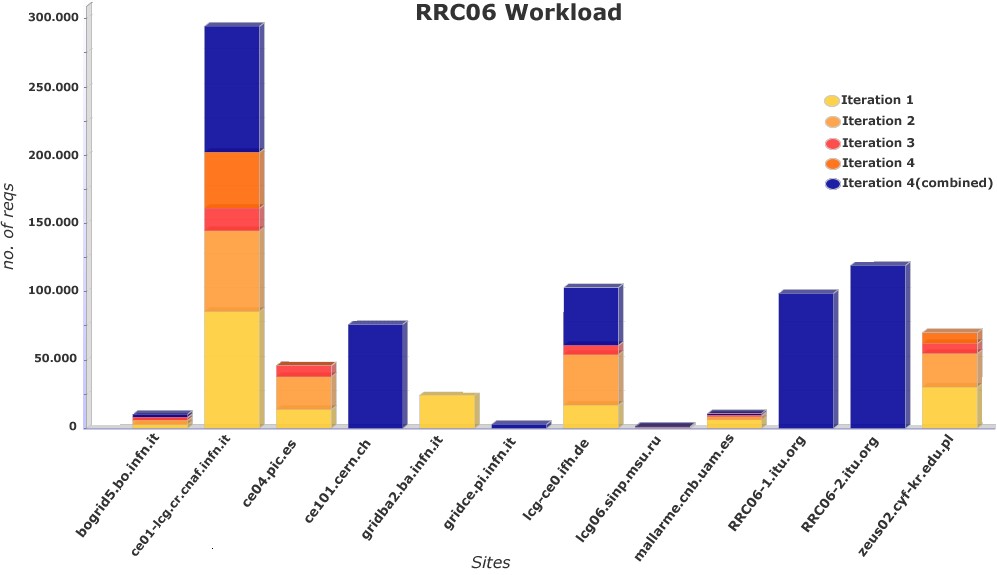}
\caption{Total workload executed in Grid and ITU clusters.}
\label{blah-itu-combined-run-barplot}
\end{figure}

Using pluggable modules, the MonALISA collector has been customized to
aggregate fine-grained data from Grid worker nodes and ITU farm nodes
to produce in real-time, higher level reports and
charts. Fig.\ref{blah-itu-combined-run-barplot} shows the total workload executed
by ITU clusters and the EGEE sites. The ITU clusters
are reported as {\tt RRC06-1.itu.org} and {\tt RRC06-2.itu.org}.

The complementary usage of Grid Unix-based and Windows-based resources
for numerical computations, required compilation of application
software on both platforms and verification of output in terms of
numerical accuracy.

%%%%%%%%%%%%%%%%%%%%%%%%%%%%%%%%%%%%%%%%%%%%%%%%%%%%%%%%
% We have a problem wth Fig. references in this section!
%%%%%%%%%%%%%%%%%%%%%%%%%%%%%%%%%%%%%%%%%%%%%%%%%%%%%%%%

%\begin{figure}
%\centering
%\includegraphics[width=10cm]{MonaLisa1}
%\label{MonaLisa2}
%\caption{System intergration via Mona Lisa}
%\end{figure}

%% file: end.tex
\section{Conclusions and Outlook}

\VISION{
conclusions:
 - success, 
 - procurement/peak usage - complementary aspect
 - dependable computing given time constraints, 
 - easy of integration of new communities
 - technical aspects: unix/windows, availability of grid sites

outlook:
 - need to a system support for the implementation of GE06
 - local/grid/clouds scenario -> added value: developing in house expertise etc. 
}

The dual system presented in this paper contributed to the success of
the RRC06 Conference which resulted in a new international treaty.

Seamless access to resources from Grid and corporate infrastructures
demonstrated in this paper may be beneficial for other user
communities. A typical use-case could include dedicated in-situ
resources for fast response and Grid resources when facing peak
demand. In such a scenario the Grid could provide a competitive
alternative to traditional procurement of resources. At RRC06 the Grid
delivered dependable peak capacity to an organization which normally
does not require a large permanent computing infrastructure. The Grid
was successfully used in a new area to provide a dependable
just-in-time service. ITU personnel needed limited support and training to adopt the Grid
technology for RRC06. This demonstrates the maturity of Grid
technology for usage in new scientific communities and technical
activities.

The outcome of RRC06 was the GE06 frequency plan which is a part of an
international agreement. Modifications to the GE06 Plan may require a
coordination examination to determine Member States potentially
affected. To bring into use a new broadcasting station a
conformity examination is required to verify that the proposed
implementation does not cause more interference than foreseen by the
GE06 Plan. Both examinations may require intensive calculations. In
addition, some Member States have already expressed the possible need
for re-planning parts of the GE06 planned bands, a process which
would imply a similar (smaller scale) approach to the one adopted at
the RRC06. 

In order to prepare for future events which may require even more
computing capabilities than the RRC06, paradigms such as Cloud
computing could be investigated, where dynamically scalable resources
are provided as a service over the Internet.  A system integrating
local, grid and cloud resources would allow Member States to submit
via an existing ITU web portal time-consuming calculation requests
and, at the back-end, to schedule and execute jobs transparently on
the integrated infrastructure. Such a pilot project could be a
continuation of the system accomplished for the RRC06 and a potential area
of future collaboration between ITU and CERN.

%Infastructure as a service (IaaS) paradigm of Cloud computing.

\section{Acknowledgments}
The authors would like first of all to thank the sites pledging resourcing for this activity. The sites are:
CNAF (Bologna, Italy) complemented by a few other sites belonging to the Italian infrastructure (Grid IT), CYFRONET (Krakow, Poland)
DESY (Hamburg and Zeuthen, Germany), MSU (MOSCOW, Russia), PIC (Barcelona, Spain).
Their willingness to share their resources was an essential condition for the success of this activity.

We would like to thank the MonALISA team (CERN/Caltech) and in
particular Iosif Legrand for support.  We would also like to thank the
ITU-IS for the ITU farm set-up and continuous support and all the
ITU-R staff involved in the data preparation for the RRC06. We are
grateful to M. Cosic and D.Botha for providing the compilation of the
compatibility analysis executable for Linux and Windows platform, and
to J. Boursy for encouraging the developing of an ITU distributed
computing system several years before it was needed at the RRC06.  A
special thank to P.N. Hai (ITU), who coordinated the overall
processing at the RRC06, for encouragement, vision and useful
feedback.  Finally a special thank goes the management of the ITU-R
and CERN IT department, in particular to T. Gavrilov and
W.~von~R\"uden (CERN IT Dept head) for encouragement and useful
discussions.

%% file: itu-paper.bbl
\begin{thebibliography}{00}

\bibitem{ITU} International Telecommunication Union, http://www.itu.int

\bibitem{RRC06FinalActs} Final Acts of the Regional Radiocommunication Conference for planning of the digital terrestrial broadcasting service in parts of 
Regions 1 and 3, in the frequency bands 174-230 MHz and 470-862 MHz (RRC-06), http://www.itu.int/ITU-R/conferences/rrc/rrc-06/index.asp

\bibitem{RRC04} Regional Radiocommunication Conference 2004 for planning of the digital terrestrial broadcasting service in parts of Regions 1 and 3, 
in the frequency bands 174-230 MHz and 470-862 MHz (RRC-04),(Geneva, Switzerland, 10 May-28 May 2004), http://www.itu.int/ITU-R/conferences/rrc/rrc-04/index.asp

\bibitem{RRC06PrepAct} Intersessional Activities (RRC-04/06),(Geneva, Switzerland, May 2004-May 2006), http://www.itu.int/ITU-R/conferences/rrc/rrc-04/intersession/index.asp

\bibitem{EGEE} Enabling Grid for E-sciencE (EGEE) home page: http://www.eu-egee.org

\bibitem{ITUConstitution} Constitution of the ITU, Chapter VII, Art. 44, ``Use of the Radio-Frequency Spectrum and of the Geostationary-Satellite and Other Satellite Orbits'',
 http://www.itu.int/net/about/basic-texts/constitution/chaptervii.aspx

\bibitem{RR4p2} International Telecommunication Union, Radio Regulations, Article 4.1

\bibitem{P1546} Recommendation ITU-R P.1546-3 Method for point-to-area predictions for terrestrial services in the frequency range 30 MHz to 3000 MHz

\bibitem{EBU} European Broadcasting Union home page, http://www.ebu.ch

\bibitem{LHC} General updated information on the LHC programme can be found on the CERN web site (http://www.cern.ch). A recent review article on the first 2 years of LHC is: Fabiola Gianotti ``Physics during the first two years of the LHC'', New J. Phys. 9 (2007) 332. DOI: 10.1088/1367-2630/9/9/332

\bibitem{OSG} Open Science Grid (OSG) Web Page, http://www.opensciencegrid.org
\bibitem{NDGF} Nordic Data Grid Facility (NDGF) Web Page, http://www.ndgf.org

\bibitem{Grid} Ian Foster and Carl Kesselman, ``The GRID: Blueprint for a New Computing Infrastructure'', Morgan Kaufmann, 1998

\bibitem{LHCC} S. Bethke et al., ``Report of the Steering Group of the LHC Computing Review'', CERN/LHC/2001-004, CERN/RRB-D 2001-3, 22 February 2001

\bibitem{LCG} LHC Computing Grid (LCG) home page: http://cern.ch/lcg

\bibitem{Monarc} Models of Networked Analysis at Regional Centers for LHC Experiments (MONARC) project home page, http://cern.ch/monarc

\bibitem{LCG-TDR} The LCG Editorial Board, ``LHC Computing Grid Technical Design Report'', LCG-TDR-001, CERN-LHCC-2005-024, June 2005

\bibitem{Ganga} J.T.Mo\'scicki et.al.: Ganga: a tool for computational-task management and easy access to Grid resources; arXiv:0902.2685v1

\bibitem{DIANE} Distributed Analysis Environment, http://cern.ch/diane

\bibitem{Monalisa} Monitoring Agents Using a Large Integrated Services (MonALISA) project home page: http://monalisa.cern.ch/monalisa.html

\bibitem{XDR} RFC1014 - XDR: External Data Representation standard, http://www.faqs.org/rfcs/rfc1014.html

\bibitem{SchedulingForResponsiveGrids}
  C.~Germain-Renaud, C.~Loomis , J.~T.~Mo{\'s}cicki and R.~Texier,
\textit{Scheduling for Responsive Grids}, 
\href{http://dx/doi.org/10.1007/s10723-007-9086-4}
{J. Grid Computing \textbf{6}, (2008) 15-27 }

\end{thebibliography}
